\documentclass[12pt]{article}
\usepackage{amssymb}
\usepackage{color,graphicx}
\usepackage[dvipsnames]{xcolor}
\usepackage{amsmath}
\usepackage{array}
\usepackage{epsfig} 
\usepackage{enumerate}
\usepackage{comment}
\usepackage[
  colorlinks=true,
  linkcolor=blue,
  citecolor=blue,
  urlcolor=blue,
]{hyperref}

\usepackage{bm}
\def\eps{{\varepsilon}}

\newcommand{\mb}{ \bm b }
\newcommand{\mB}{ \bm B }
\newcommand{\mE}{ \bm E }
\newcommand{\mj}{ \bm j }
\newcommand{\mk}{ \bm k }
\newcommand{\mv}{ \bm v }
\newcommand{\mx}{ \bm x }

\def\eps{\varepsilon}
\def\boldnabla{{\bm \nabla}}

\newcommand{\keywords}[1]{\par\noindent\textbf{Keywords:} #1\par}

\usepackage{xcolor}
\usepackage{comment}

\allowdisplaybreaks
\allowbreak

\begin{document}

\title{Magnetohydrodynamics in turbulent dynamo regime: the stability problem}

\author{ Michal Hnati\v{c}$^{1,2,3}$,
  Tom\'{a}\v{s} Lu\v{c}ivjansk\'{y}$^{2}$,
  Luk\'a\v{s} Mi\v{z}i\v{s}in$^{1}$,\\
  Yurii Molotkov$^{1}$, and Andrei Ovsiannikov$^{1}$.
  }

\maketitle              
\mbox{ }\\
$^1$ Joint Institute for Nuclear Research, Dubna, Moscow Region, Russia \\
$^2$ Institute of Physics, Faculty of Science, Šafárik University, Košice, Slovakia\\
$^3$ Institute of Experimental Physics, Slovak Academy of Sciences, Košice, Slovakia\\

\begin{abstract}
This paper investigates stochastic solenoidal magnetohydrodynamics 
within the field-theoretic Martin--Siggia--Rose--De Dominicis--Janssen formalism, with a specific focus on the stability of the system when spatial mirror (parity) symmetry is explicitly broken.
Under helical forcing, the one-particle-irreducible magnetic response function already at one loop contains a curl-type contribution that dominates the bare resistive term in the infrared limit, leading to exponential instability of the trivial state $\langle \mb \rangle = {\bm 0}$.
We re-examine a stabilization mechanism proposed in [L. T. Adzhemyan, et al., Theor. Math. Phys. \textbf{72}, 940--950 (1987)], in which the system evolves into a phase with a dynamically spontaneously broken rotational symmetry and a generated mean magnetic field $\langle \mb \rangle = \mB_0$.
By deriving a self-consistency condition for $\mB_0$, we show that for any physically admissible (infrared) form of the pumping function, the model admits only a singular solution.
We illustrate this with the standard power-law and ``massive'' pumping functions.
We further show that previous claims of a finite $\mB_0$ arose from an inconsistent truncation of asymptotic expansions.
We argue that a consistent physical resolution requires including a bare curl term in the stochastic induction equation, which naturally arises from a parity-violating modification of Ohm’s law.
With this modification, stabilization of the system by spontaneous symmetry breaking becomes a viable field-theoretic description of large-scale mean-field generation (turbulent dynamo) in helical turbulent magnetohydrodynamics.
\vspace{0.3cm}

\end{abstract}

\section{Introduction}


Magnetohydrodynamics (MHD) of turbulent conducting media is a classical problem in theoretical physics and applied mathematics, with applications ranging from astrophysics and geophysics to plasma physics, as well as laboratory and industrial flows.
For a representative overview and further references, see \cite{Hnatic2025}.
In the regime of fully developed turbulence, the dynamics of such systems is essentially governed by the nonlinear hydrodynamic and electromagnetic interactions.
Their statistical properties in this regime can be described within a model of statistically stationary (forced) turbulence supplemented by stochastic sources that model energy injection at large scales.

A crucial point is that the absence of spatial mirror (parity) symmetry is the rule rather than the exception for realistic MHD systems.
Parity can be broken by geometry, rotation, stratification, or other mechanisms, and this typically enables additional effects beyond those present in parity-invariant turbulence.
In such parity-breaking systems (often referred to as helical or gyrotropic in the older terminology), specific phenomena arise, including the well-known $\alpha$-effect and related contributions that play an essential role in magnetic self-organization and dynamo processes \cite{Krause_MFEandDynamo, Vainshtein_TurbulentDynamo, Moffat_MFGen, Davidson_InttoMHD}.

In this work, we investigate stochastic solenoidal MHD within the field-theoretic Martin--Siggia--Rose--De Dominicis--Janssen (MSRDJ) formalism \cite{MSR1973, Janssen1976, DeDominicisPeliti1978, Tauber2014, Vasiliev2004}.
This powerful approach reformulates the original stochastic problem in terms of a statistical field theory with a well-defined action functional and the corresponding diagrammatic (Feynman) rules.
The form of energy injection (the pumping function) is chosen in a way standard for the renormalization-group (RG) treatment of stationary turbulence, namely, such that its Fourier kernel has a power-law asymptotic behavior at large momenta \cite{Vasiliev2004}.
With this choice, ultraviolet (UV) divergences appear in the theory; they are then systematically removed by UV renormalization, which makes it feasible to apply RG techniques to the calculation of universal quantities.

The central problem addressed in this paper arises when parity symmetry is explicitly broken.
In our model, parity breaking enters through the helical (parity-odd) part of the forcing correlator.
This fact leads to the appearance, already at one-loop order, of an infrared-relevant contribution linear in the external momentum $\mk$ and proportional to $i\epsilon_{ijm}k_m$, in the one-particle-irreducible (1PI) magnetic response function \cite{Pouquet1978} (see also \cite{Hnatic2025, Hnatic87}).
Importantly, the coefficient of this curl (rotor) term is UV linearly divergent, i.e., it grows linearly with the UV cutoff $\Lambda$.
As a result, in the infrared limit $\mk \to 0$ the curl term dominates over the dissipative bare resistive contribution $\propto k^2$, rendering the trivial state $\langle \mb\rangle = \bm 0$ exponentially unstable.

The main goal of this paper is to clarify a subtle point related to the stabilization mechanism proposed in \cite{Hnatic87} for such helical MHD systems.
The idea is that the instability is removed by means of a large-scale magnetic field arising in it during the process of dynamical spontaneous breaking of rotational symmetry.\footnote{It should be noted that in \cite{Hnatic87} and \cite{Hnatic2025}, another, more ``adjustable'' and formal mechanism for eliminating instability was noted and investigated, leading to a simplified theory, the so-called kinematic theory of dynamo \cite{Krause_MFEandDynamo, Vainshtein_TurbulentDynamo}.
See \cite{Hnatic2025} for details.}
This mechanism was investigated in detail in \cite{Hnatic2025}; however, due to space constraints, a sensitive technical issue was not discussed there in sufficient detail -- namely, the correct determination of the magnitude of the stabilizing field in the first (one-loop) approximation, on which the subsequent two-loop construction in \cite{Hnatic2025} is based.
The purpose of the present paper is to fill this gap.

Our main result is a detailed re-analysis and a consistent formulation of the stabilization mechanism according to which the system enters a broken-symmetry phase with a spontaneously generated mean magnetic field $\langle \mb\rangle = \mB_0$.
The magnitude of $\mB_0$ is assumed to be fixed self-consistently to cancel the unstable curl contribution.
At the same time, as already noted in \cite{Hnatic87}, this cannot be implemented by the standard route of passing to the generating functional of 1PI Green functions (the effective action or ``free energy'') $\Gamma$ and analyzing its extrema, since for models of our type the extremum condition for $\Gamma$ does not fix spatio-temporally homogeneous configurations of the averages.
Therefore, assuming the existence of a vacuum state $\mB_0$ in the system, we shift to its vicinity at the level of the generating functional of connected (rather than 1PI) Green functions and work directly in the broken-symmetry phase.
We then study the one-loop contribution to the 1PI function $\Gamma^{b'b}$ that is linear in the external momentum (the curl term).
Since in the MSRDJ framework $\big(\Gamma^{b'b}\big)^{-1}$ is the magnetic response function, the cancellation of the unstable curl contribution yields a self-consistency condition that determines the modulus $B_0 \equiv |\mB_0|$.
The direction of $\mB_0$ remains unfixed, as usual in systems with spontaneous symmetry breaking.

We also emphasize that although the value of $\mB_0$ depends on the specific form of the pumping function (it is a nonuniversal amplitude from the RG viewpoint), the stabilization mechanism itself should, in principle, be compatible with any physically admissible choice of pumping.
For the standard forms of the pumping function, both purely power-law and massive, we show that the self-consistency condition admits only a singular limit solution $B_0 \to \infty$, in contrast to the finite result reported in \cite{Hnatic87}.
This indicates that, for natural choices of pumping, the homogeneous-field stabilization mechanism does not yield a finite stabilizing magnetic field without additional physical input or an explicit regularization.
This is precisely the subtle but crucial point that is clarified in this paper.

The finite value of $B_0$ reported in Ref.~\cite{Hnatic87}, which contrasts with our findings, arises from a computational inconsistency.
The self-consistency equation for $B_0$ is given by the sum of all one-loop diagrams contributing to the magnetic response function and can be written schematically as
\begin{equation}
\label{eq:self-consistency_equation_for_B0}
\mathrm{const}_1 \Lambda^{1 - 2 \epsilon} - \mathrm{const}_2 \, B_0^{1 - 2 \epsilon} \int_0^{\Lambda/B_0}\mbox{d} k \,f(k) = 0,
\end{equation}
where $\mathrm{const}_{1,2}$ and the function $f(k)$ depend on the remaining parameters of the model (in particular, on $\epsilon > 0$ and the kinematic viscosity).
Introducing a dimensionless parameter $c$ via $B_0 = c\sqrt{u_0}\,\nu_0 \Lambda$ and analyzing \eqref{eq:self-consistency_equation_for_B0} in the asymptotic regime $\Lambda \to \infty$, one immediately observes that the upper limit of the (formally UV-finite) integral remains finite, rather than tending to infinity -- as it would in the computation of a genuinely UV-finite quantity independent of $\Lambda$ -- precisely because $B_0 \sim \Lambda$.
The inaccuracy in \cite{Hnatic87} appears to be that the authors effectively replaced the finite upper limit in the UV-finite integral in \eqref{eq:self-consistency_equation_for_B0} by infinity at an early stage and only then solved for $B_0$.
As noted in \cite{Hnatic2025}, such a step is natural in a completely renormalized formulation, and it reproduces the finite answer of \cite{Hnatic87} when expressed in terms of a UV-finite analogue of $\Lambda$.
However, for the renormalization procedure to be self-consistent at all stages, the original bare equation \eqref{eq:self-consistency_equation_for_B0} must also be solved, and, as we show, it admits only the singular solution $c \to \infty$.

On this basis, we discuss the possible resolution of the problem.
A natural resolution is to recognize that once mirror symmetry is absent, a curl term with a finite coefficient is allowed in the equations of motion and may already be present there (with some UV-finite coefficient) from the outset.
This can be viewed as a ``seed'' curl contribution with a small coefficient $h_{\mathrm{fin}}$, serving as an effective regularization.
This plays the role of a small symmetry-allowed seed (a mass-like regulator), somewhat analogous in spirit to introducing a regulator mass in quantum field theory.
We demonstrate that for arbitrarily small $h_{\mathrm{fin}}$ of the appropriate sign, a finite solution for $B_0$ exists (and can even be of modest magnitude for very small $h_{\mathrm{fin}}$).
One may argue that such a finite seed is generated during the onset of turbulence, for instance, as a manifestation of the mean turbulent electromotive force.
From this perspective, our field-theoretic description should be regarded as an effective framework for the steady developed turbulent state and thus applies once the system has already reached a stable regime in which the initial instability has been eliminated -- potentially via the emergence of the spontaneous large-scale field $\mB_0$.
Accordingly, the value of $B_0$ is fixed by physical processes operating before the establishment of the stationary turbulent state; indeed, achieving a steady state necessarily requires that the instability be removed first.

Summarizing, a consistent resolution of the instability by means of spontaneous symmetry breaking requires acknowledging that the total curl contribution contains, besides the diagrammatically generated part in the magnetic response function, a finite seed component carried into the stationary state from the pre-stationary dynamics.
In the presence of such a seed, the self-consistency equation for $B_0$ may admit a finite solution: for nonzero $B_0$, the finite seed curl contribution together with the curl term generated by diagrams of magnetic response can cancel each other, so that the net curl contribution vanishes, and the magnetic response remains stable for $\mk\to 0$.
As a result, for a given UV cutoff $\Lambda$, the self-consistency equation for $B_0$ can indeed be solved at a finite value.

The paper is organized as follows.
In Section~\ref{sec:QFT}, we derive the MSRDJ action for the stochastic MHD equations and specify the statistics of the only external (Gaussian) noise via its pair correlator, whose Fourier kernel is the pumping function.
Particular emphasis is placed on the commonly used purely power-law and ``massive'' forms of this kernel.
In Section~\ref{sec:Curl_terms}, we briefly summarize the findings of \cite{Hnatic2025} concerning the origin of the curl-type linear divergences and the resulting instability.
In Section~\ref{sec:Pumping_function}, we discuss in detail the one-loop self-consistency equation that determines the magnitude of the stabilizing field $B_0$, possible regularizations leading to a finite $B_0$, and the physical motivation for introducing them.
Finally, in Section~\ref{sec:concl} we summarize our conclusions and outline directions for future research, including an extended stochastic MHD model with explicit magnetic and mixed noise terms and an RG analysis of the composite operator $\boldsymbol{v}\times\boldsymbol{b}$, whose mean is associated with the $\alpha$-effect.

\section{Field-theoretic formulation of the model}
\label{sec:QFT}

The theoretical framework of non-relativistic three-dimensional stochastic magnetohydrodynamics relies on two fundamental equations: the
Navier--Stokes equation for an incompressible fluid driven by a Gaussian helical forcing, and an induction equation for the magnetic field derived from Maxwell’s equations.
In Alfv\'en units, the system reads
\begin{equation}
\label{Eq:MHD_Eq}
\begin{split}
&\partial_t\boldsymbol{v} + (\boldsymbol{v} \cdot \boldsymbol{\nabla})\boldsymbol{v} = - \boldnabla \big(p + \boldsymbol{b}^2/2\big) + \nu_0 \boldnabla^2 \boldsymbol{v} + 
(\boldsymbol{b} \cdot \boldnabla) \boldsymbol{b} + \boldsymbol{f}^{v}, \\
&\partial_t\boldsymbol{b} + (\boldsymbol{v} \cdot \boldsymbol{\nabla})\boldsymbol{b} = \kappa_0 
\boldnabla^2 \boldsymbol{b}
+  (\boldsymbol{b} \cdot \boldnabla) \boldsymbol{v}, \\
&\qquad\boldnabla \cdot \boldsymbol{v} = 0, \qquad\qquad
\boldnabla \cdot \boldsymbol{b} = 0, \qquad\qquad
\boldnabla \cdot \boldsymbol{f}^{v} = 0.
\end{split}
\end{equation}
Here $\bm v(\bm x,t)$ denotes the fluctuating Eulerian velocity field, while
$\mb(\mx,t) = \mB(\mx,t)/\sqrt{4\pi\rho_0}$ is the magnetic field in Alfv\'en units; in general, $\mb$ may include a deterministic (mean) component in addition to fluctuations.
The scalar field $p(\mx,t)$ is the hydrodynamic pressure enforcing incompressibility, and $\bm f^{v}(\mx,t)$ is an external random force.
We use $\partial_t\equiv\partial/\partial t$, $\boldnabla$ for the spatial gradient, and $\boldnabla^2$ for the Laplacian.
The parameters $\nu_0$ and $\kappa_0 = u_0\nu_0$ are the kinematic viscosity and the magnetic diffusivity, respectively, where $u_0 = \mathrm{Pr}_m^{-1}$, and $\mathrm{Pr}_m$ is the magnetic Prandtl number.
Note that here, for simplicity, we drive the system mechanically (in the velocity equation) and do not introduce an independent magnetic forcing; magnetic fluctuations are generated dynamically via the induction term.

The second equation in \eqref{Eq:MHD_Eq} follows from Maxwell's equations
$\boldnabla\cdot\mB = 0$ and $\boldnabla\times\mE + c^{-1}\partial_t\mB = 0$, together with Amp\`ere's law in the quasi-static limit, $\boldnabla\times\mB = 4\pi\mj/c$ (displacement current neglected), and the simplest form of Ohm's law for a moving conductor,
$\mj = \sigma_0\big(\mE + c^{-1}[\mv\times\mB]\big)$.
Throughout, $c$ is the speed of light and Gaussian units are used.

To maintain translational invariance in time, we impose vanishing boundary conditions at spatial infinity and asymptotic initial conditions in the remote past,
\begin{equation}\label{eq:MHD_initial_and_boundary_cond}
    \lim_{\|\bm{x}\| \rightarrow \infty} \boldsymbol{v}(\bm{x}, t) = 0, \qquad  \lim_{\|\boldsymbol{x}\| \rightarrow \infty} \boldsymbol{b}(\bm{x}, t) = 0, \qquad
   \lim_{t \rightarrow -\infty} \boldsymbol{v}(\bm{x}, t) = 0, \qquad \lim_{t \rightarrow -\infty} \boldsymbol{b}(\bm x, t) = 0.
\end{equation}
These asymptotic conditions select the unique statistically stationary state generated by the forcing.

The stochastic problem \eqref{Eq:MHD_Eq}, together with the conditions \eqref{eq:MHD_initial_and_boundary_cond}, can be recast into a field-theoretic form by virtue of the following MSRDJ action:
\begin{align}\label{eq:MHD_action_without_shift}
\mathcal{S}_0 = \frac{1}{2}\boldsymbol{v'} \mathfrak{D}^v \boldsymbol{v'} &+ \boldsymbol{v'}\cdot \left[-\partial_t\boldsymbol{v} - (\boldsymbol{v} \cdot \boldsymbol{\nabla})\boldsymbol{v} + {\nu}_0 \boldsymbol{\nabla}^2 \boldsymbol{v} + (\boldsymbol{b} \cdot \boldsymbol{\nabla}) \boldsymbol{b}\right]  \notag \\
&+\boldsymbol{b'}\cdot\left[-\partial_t\boldsymbol{b} - (\boldsymbol{v} \cdot \boldsymbol{\nabla})\boldsymbol{b} + u_0 {\nu}_0 \boldsymbol{\nabla}^2 \boldsymbol{b} + (\boldsymbol{b} \cdot \boldsymbol{\nabla}) \boldsymbol{v}\right].
\end{align}
Here $\bm v'$ and $\bm b'$ are the auxiliary (response) fields arising in the MSRDJ construction \cite{Vasiliev2004}; they satisfy the same boundary/asymptotic conditions as $\bm v$ and $\bm b$.
The pressure term is eliminated by projecting the Navier--Stokes equation onto the transverse (solenoidal) sector; equivalently, it drops out upon contraction with the transverse response field $\bm v'$.
The subscript ``0'' indicates bare (unrenormalized) quantities.
To avoid cluttering the notation, we henceforth suppress the subscript on fields when this does not cause ambiguity.
Integration over space--time is understood, and explicit integral signs are omitted.

The kernel $\mathfrak{D}^v$ is determined by the statistics of the random force $\bm f^v$.
We assume $\bm f^v$ to be a simplest stationary random process, i.e., a white-in-time Gaussian random field with zero mean, completely specified by its two-point correlation function $\mathfrak{D}^v$.
The correlator 
$\mathfrak{D}^v$ is then defined as:
\begin{equation}
\label{eq:force_correlator}
\mathfrak{D}_{i j}^v(\bm{x} - \bm{x}', t - t') \equiv \langle f_i^{v} (\bm{x}, t) f_j^{v} (\bm{x}', t') \rangle = \delta(t - t') \int \frac{\mbox{d}^d k}{(2 \pi)^d} \mathbb{D}_{i j} (\bm{k}) \mathrm{e}^{i \bm{k} \cdot (\bm{x} - \bm{x}')},
\end{equation}
where $\langle\ldots\rangle$ denotes averaging over the distribution of $\bm f^v$, and $\mathbb{D}_{ij}(\bm{k})$ is referred to as the pumping (forcing) spectrum.
The $\delta(t - t')$ time dependence is a standard idealization in RG turbulence models, which preserves Galilean invariance and simplifies the analysis.
In the context of homogeneous, fully developed MHD turbulence, we take
\begin{equation}\label{eq:helical_pumping}
\mathbb{D}_{ij}(\mk) = \left(\mathbb{P}_{ij}(\mk) + i\rho \epsilon_{ijm}\frac{k_{m}}{k}\right)N(k),
\end{equation}
with $\mathbb{P}_{ij}(\bm{k}) = \delta_{ij} - k_i k_j/k^2$ the transverse projector, $k = |\bm{k}|$, $\epsilon_{ijm}$ is the totally antisymmetric Levi-Civita tensor.
The parameter $\rho$ quantifies mirror-symmetry breaking (helicity) in the forcing and is restricted by realizability (positive definiteness of the correlator).

Regarding the choice of the random force correlator, the following should be emphasized.
It is convenient to distinguish two broad classes of stochastic dynamical models within the MSRDJ formalism.
In the first class, the long-time dynamics relaxes to an equilibrium state described by a static (time-independent) action; in this case, the random-force correlator is not arbitrary but is uniquely fixed by the dynamical-to-static correspondence (see, e.g., \cite{Vasiliev2004}).
In the second class, the system remains permanently out of equilibrium, and the correlator of the driving noise constitutes part of the model definition.
Fully developed turbulence, including stochastic magnetohydrodynamics, belongs to the latter class: the system approaches a statistically stationary but intrinsically non-equilibrium turbulent state (see, e.g., \cite{Frisch}).
Thus, the choice of \eqref{eq:force_correlator}--or, in turbulence terminology, the choice of the pumping $N(k)$--is a crucial ingredient in specifying a model of MHD turbulence. 
Note that, in the case of ordinary hydrodynamic turbulence, there exists a result \cite{Adzhemyan1998} that makes it possible to reconstruct a physically realistic form of the pumping, at least on the energy-injection scales and throughout the inertial range.

Within the MSRDJ approach, the standard formulation of homogeneous fully developed MHD turbulence is aimed at universal characteristics: scaling (critical) exponents--equivalently, the spectral exponents of kinetic and magnetic energies--and the corresponding normalized scaling functions.
Universality here means insensitivity to microscopic details and, in particular, to the detailed shape of the pumping, provided it is physically admissible.
Since the forcing represents energy injection by large-scale eddy motions, the mean power input must be dominated by the small-wavenumber region.
Accordingly, one often adopts the form
\begin{equation}
\label{eq:mass_pump}
N(k) = g_{0}\nu_0^3 k^{4 - d}(k^{2} + m^{2})^{-\epsilon},
\end{equation}
where $g_0 \sim \Lambda^{2\epsilon}$ plays the role of coupling constant, the $\Lambda$ is an ultraviolet momentum scale (e.g., the maximum of inverse viscous and dissipative lengths), $d$ is the spatial dimension (in our case $d = 3$), $m$ is an infrared scale (inverse integral length), and $\epsilon > 0$ is the $\epsilon$-expansion parameter in RG turbulence theory.
It plays the role analogous to the $4 - d$ expansion parameter in the theory of critical phenomena.
The physical infrared-dominated forcing becomes relevant for $\epsilon\ge 2$.
For more on this formulation and on the construction of the $\epsilon$-expansion in turbulent models, see monographs \cite{Adzhemyan_RGinFullDevTurb, Vasiliev2004}.

In many practical applications, one further employs a pure power-law pumping with a sharp infrared cutoff
\begin{equation}
\label{eq:power_pump}
N(k) = g_{0}\nu_0^3 k^{4 - d - 2\epsilon}\theta(k - m),
\end{equation}
where $\theta(x)$ is the Heaviside step function.

For either choice, the resulting field theory requires ultraviolet renormalization: Feynman diagrams involving the noise correlator develop UV divergences.
After their subtraction, the renormalization-group analysis yields universal observables as expansions in the formal parameter $\epsilon$.

\section{Remarks on UV renormalization, curl terms, and the turbulent-dynamo regime}
\label{sec:Curl_terms}

In brief, ultraviolet renormalization is usually carried out in two stages: a $\Lambda$--renormalization followed by an $\epsilon$--renormalization.
At the first stage (typical for multiplicatively renormalizable models in the theory of critical phenomena), power-law divergences in $\Lambda$ are absorbed by a renormalization of the infrared-relevant parameters of the action.
The second stage is the standard $\epsilon$--renormalization.
At this level, for our parity-violating model~\eqref{eq:MHD_action_without_shift}, the renormalization program was recently carried out to two-loop order in~\cite{Hnatic2024}.
A complete two-loop renormalization of the model~\eqref{eq:MHD_action_without_shift}, including the nontrivial $\Lambda$--renormalization required in the present case, is discussed in full detail in~\cite{Hnatic2025}; here we only summarize the main conclusions.

For $d > 2$, all superficial UV divergences in~\eqref{eq:MHD_action_without_shift} occur only in the one-particle-irreducible (1PI) functions
$\Gamma^{v'v}$, $\Gamma^{b'b}$, and $\Gamma^{v'bb}$.
One then finds that, in addition to the usual logarithmic divergences, the 1PI function $\Gamma^{b'b}$ also develops a divergence linear in the UV cutoff $\Lambda$.
It was also proven in~\cite{Hnatic2025} (in a general form) that such contributions are forbidden in $\Gamma^{v'v}$ and $\Gamma^{v'bb}$ by the symmetry properties of the corresponding interaction vertices.

Besides the renormalization issue, the theory~\eqref{eq:MHD_action_without_shift} faces an additional difficulty.
In the stochastic dynamics~\eqref{Eq:MHD_Eq} formulated within the MSRDJ formalism, the (local) stability of the system is determined by the response functions
$\langle \boldsymbol{v}' \otimes \boldsymbol{v}\rangle = \big(\Gamma^{v'v}\big)^{-1}$
and
$\langle \boldsymbol{b}' \otimes \boldsymbol{b}\rangle = \big(\Gamma^{b'b}\big)^{-1}$
\cite{Hnatic87}.
At tree level, stability is ensured by the correct (dissipative) signs of the seed terms
$\sim \nu_0 k^{2}$ and $\sim \nu_0 u_0 k^{2}$.
However, at one-loop order $\Gamma^{b'b}$ contains a contribution linear in $\Lambda$ and, consequently, linear in the external momentum $k$, with tensor structure $\epsilon_{ijm}k_m$.
Therefore, in the infrared (small-$k$) asymptotic regime, this term is more relevant than the seed term $\nu_0 k^{2}$ and hence dominates.
As a result, the system becomes unstable, which calls into question the physical consistency of the original model.
The contributions $\sim \epsilon_{ijm}k_m\Lambda$ that arise in diagrams for the 1PI function $\Gamma^{b'b}$ correspond to an action term proportional to $\sim \operatorname{curl}\,\boldsymbol{b}$; for this reason, they are referred to as curl terms---the issue addressed below.

The loss of stability and the impact of curl terms on the linearized MHD equations were analyzed in~\cite{Hnatic87, Hnatic2025}, where it was concluded that the only logically consistent resolution is to eliminate the curl terms from the theory to all orders in perturbation theory.
In essence, this conclusion is reached because the rotor terms are linear in the UV cutoff $\Lambda$ and renormalize an infrared-relevant parameter with the same tensorial structure---a parameter that is absent in~\eqref{eq:MHD_action_without_shift} and, moreover, whose formal inclusion into the model immediately renders it unstable.
In~\cite{Hnatic2025}, two possible mechanisms were explored.
The first effectively amounts to an exact cancellation of the diagrammatically generated rotor contribution by a suitably tuned seed counterterm.
Equivalently (and more simply stated), one may adopt the formal rules of analytic regularization\footnote{We recall that in the RG theory of turbulence the parameter $\epsilon$ is not related to the spatial dimension $d$. Technically, however, this regularization is analogous to the familiar dimensional regularization.} in the parameter $\epsilon$
under which such infrared-relevant objects are set to zero and hence ignored.
This prescription is justified, for example, in the theory of critical phenomena: emergent $\Lambda$--terms (as in the $\varphi^{4}$ theory, the universality class of the Ising model) are handled via a so-called $\Lambda$--renormalization, where the $\Lambda$--term is interpreted as a shift of the critical temperature.
Although this shift can be computed within the model, within the ideology of critical phenomena, it is a non-universal quantity and therefore lies outside the scope of the theory of critical behavior.

In the present paper, we focus on the second, more fundamental scenario, termed in~\cite{Hnatic2025} the turbulent-dynamo regime of the model~\eqref{eq:MHD_action_without_shift}.
As noted above, rotor terms render the trivial state $\langle \boldsymbol{b} \rangle = {\bm 0}$ unstable.
Hence, the central idea---discussed in greater detail in~\cite{Hnatic2025}---is that the original state with $\langle \boldsymbol{b} \rangle = {\bm 0}$ is unstable, and the system evolves into a state with a spontaneously generated homogeneous magnetic field $\boldsymbol{B}_0$, i.e.,
$\langle \boldsymbol{b} \rangle = \boldsymbol{B}_0$,
which, according to~\cite{Hnatic87}, should stabilize the dynamics.
In~\cite{Hnatic87} it was proposed to fix the magnitude of $\boldsymbol{B}_0$ self-consistently to cancel the curl term exactly, and a corresponding one-loop calculation was performed.
However, as already noted in the Introduction, the computation of the stabilizing field magnitude $\boldsymbol{B}_0$ contained an error, leading (at finite $\Lambda$) to a finite result $|\boldsymbol{B}_0| \sim \Lambda$.
This seemingly plausible outcome is, unfortunately, incorrect: as we show in the next section, a careful cancellation of the one-loop rotor terms without any additional regularization leads to $|\boldsymbol{B}_0| = c\sqrt{u_0}\,\nu_0 \Lambda$ with $c = \infty$.
Thus, the dimensionless coefficient $c$ in this ansatz, written out from dimensionless considerations, diverges, and $\boldsymbol{B}_0$ self-consistency collapses. 

\section{Stabilization mechanism and pumping function}
\label{sec:Pumping_function}

In the turbulent dynamo regime, the unstable $\mathbf{SO}(3)$-symmetric state with $\langle \bm b \rangle = \bm 0$ evolves into an $\mathbf{SO}(2)$-symmetric phase characterized by a uniform mean magnetic field $\langle \bm b \rangle = \bm B_0$.
In this spontaneously broken phase, we shift the magnetic field as $\bm b \to \bm b + \bm B_0$, so that the shifted field $\bm b$ represents purely fluctuations with $\langle \bm b \rangle = {\bm 0}$.
Expanding the action about the new stationary state, one finds that for \eqref{eq:MHD_action_without_shift} the shift modifies only the quadratic part (i.e., the propagator matrix).
Explicitly, the action functional in the dynamo regime reads:
\begin{equation}
\label{eq:MHD_action_shift}
\begin{split}
\mathcal{S}^{B}_{0}&=\mathcal{S}_0+\bm{v}'\cdot(\bm{B}_0\cdot\bm{\nabla})\bm{b}+\bm{b}'\cdot(\bm{B}_{0}\cdot\bm{\nabla})\bm{v}.
\end{split}
\end{equation}
The diagrammatic rules and all details of the computational techniques (in many ways more complex than those usually encountered in turbulence theory) for the model \eqref{eq:MHD_action_shift} were provided in \cite{Hnatic2025} and are omitted here for brevity.

At first order in perturbation theory, the one-particle-irreducible function $\Gamma^{b'b}$ receives contributions from the following four diagrams:
\begin{align}\label{eq:one-loop_diagrams_after_shift}
      \Sigma_1^{b'b} = ~\raisebox{0ex}{\includegraphics[width=2.8cm]{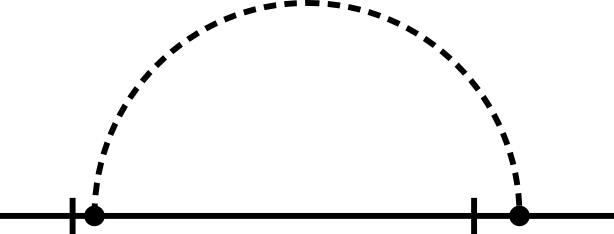}} ~+~ \raisebox{0ex}{\includegraphics[width=2.8cm]{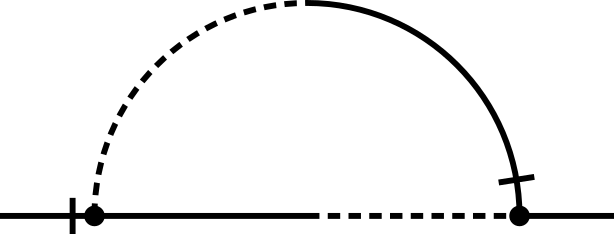}} + ~ \raisebox{0ex}{\includegraphics[width=2.8cm]{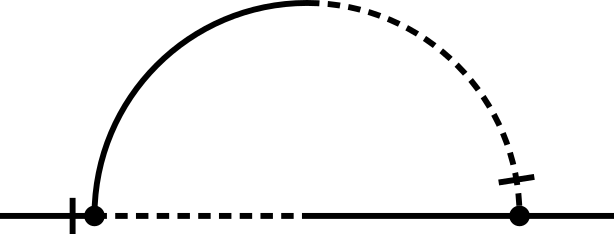}} ~+~ \raisebox{0ex}{\includegraphics[width=2.8cm]{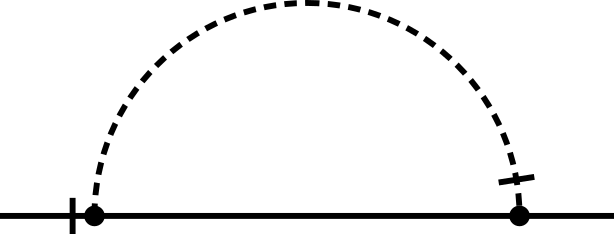}}.
\end{align}
Here $\Sigma_1^{b'b}$ denotes the one-loop contribution to $\Gamma^{b'b}$; dotted lines correspond to the velocity field, and solid lines to the magnetic field.
A slash on a line denotes the corresponding response field. 
For further details, we refer the interested reader to our previous paper \cite{Hnatic2025}.

In what follows, we retain in $\Sigma_1^{b'b}$ only the term linear in the external momentum $\boldsymbol{p}$, i.e., the contribution proportional to $i\epsilon_{i j m} p_m$ (the only linear structure allowed by symmetry), and denote it by $\Sigma_{1\, \mathrm{curl}}^{b'b}$.
After the frequency integration, which in this case can be carried out based on the arguments given by the Hermite-Biehler theorem (see \cite{Hnatic2025} for details), the remaining momentum integral takes the form
\begin{equation}
\label{eq:ceq}
\Sigma_{1\, \mathrm{curl}}^{b'b} = -\frac{i g_0 u_0 \nu_0}{2 \pi ^2 (u_0 + 1)} \left(\frac{B_0}{\nu_0}\right)^{-3}\int\limits_{0}^{\Lambda} \mbox{d}k\, k \left(\frac{B_0}{\nu_0}-k \sqrt{u_0} \cot ^{-1}\left(\frac{k\nu_0 \sqrt{u_0}}{B_0}\right)\right)N(k).
\end{equation}
Here, we keep the pumping function $N(k)$ in an arbitrary form.
The function $N(k)$ may depend on the parameters $\epsilon$ and $m$, as in Eqs.~\eqref{eq:mass_pump} and \eqref{eq:power_pump}, and possibly 
on other model parameters.
For a given choice of $N(k)$, one can then examine for which parameter values the condition $\Sigma_{1\, \mathrm{curl}}^{b'b} = 0$ can be satisfied.
In what follows, we impose the ansatz $B = c\, \sqrt{u_0}\,\nu_0\Lambda$, which allows one to analyze \eqref{eq:ceq} in terms of the dimensionless parameter $c$.

We begin with the simplest purely power-law pumping defined in Eq.~\eqref{eq:power_pump}. 
For this choice, the integral in \eqref{eq:ceq} can be completely evaluated and gives
\begin{equation}
\label{eq:pure_pump_sigma}
\begin{split}
&\Sigma_{1\, \mathrm{curl}}^{b'b} =\frac{i g_0 (\Lambda\nu_0)^{1 - 2\epsilon}}{8 \pi^2 (u_0 + 1)}
\frac{\big(\pi  (3-2 \epsilon ) c^{4 - 2 \epsilon } \sec (\pi \epsilon) + 4 c (\epsilon - 2) + 2 (3 - 2 \epsilon ) \tan^{-1}(c) + 2 c J_0(c,\epsilon)\big)}{c^3 (\epsilon -2) (2 \epsilon -3)}.
\end{split}
\end{equation}
Here $J_0(c,\epsilon)$ is defined by
\begin{equation}
J_0(c,\epsilon) \equiv \begin{cases}
 _2F_1\left(1,\epsilon -\frac{3}{2};\, \epsilon -\frac{1}{2};\, -c^2\right), & \qquad 0 < c < 1,\\
_2F_1\left(1,1;\, \epsilon -\frac{1}{2}; \,\,\frac{c^2}{(c^2 + 1)}\right)(c^2 + 1)^{-1}
, & \qquad c > 1.
\end{cases}
\end{equation}
where ${}_pF_q$ denotes the generalized hypergeometric function:
\begin{equation*}
{}_pF_q(a_1,\dots,a_p;\, b_1,\dots,b_q;\, z) = \sum_{n=0}^{\infty}
\frac{(a_1)_n \cdots (a_p)_n}{(b_1)_n \cdots (b_q)_n}
\frac{z^n}{n!}, \qquad (a)_n \equiv \frac{\Gamma(a+n)}{\Gamma(a)},\qquad (a)_0 = 1.
\end{equation*}

\begin{figure}[h]
  \centering
  \includegraphics[width=0.75\textwidth]{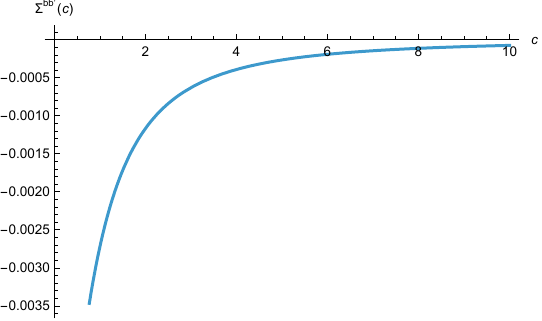}
  \caption{Graphical representation of the dimensionless function $F(c, 0)$ for the case of pumping of the type \eqref{eq:power_pump} and its tendency to asymptote $F(c, 0) = 0$.}
  \label{fig:graph1}
\end{figure}

As is well known, the applicability of the purely power-law pumping \eqref{eq:power_pump} is restricted as $\epsilon \to 2^-$ \cite{Vasiliev2004}.
Our analysis shows that for all $\epsilon \in [0;2)$, except for isolated singular points, $\Sigma_{1\, \mathrm{curl}}^{b'b}$ exhibits the same qualitative behavior, and the equation $\Sigma_{1\, \mathrm{curl}}^{b'b} = 0$ is satisfied only in the singular limit $c \to \infty$
For illustration purposes, we plot the dimensionless function $F(c, \epsilon) \equiv \Sigma_{1\, \mathrm{curl}}^{b'b}/(i g_0 (\nu_0\Lambda)^{1 - 2 \epsilon} (u_0 + 1))$ for the RG-relevant case $\epsilon = 0$ (see Fig. \ref{fig:graph1}):
\begin{equation}
\label{eq:pure_c_eq}
F(c, 0) = \frac{\pi  c^4-2 \left(c^4-1\right) \tan ^{-1}(c)-2 \left(c^3+c\right)}{32 \pi ^3 c^3}.
\end{equation}

As for the ``physical'' limit $\epsilon \to 2^-$, the situation is more delicate.
The reason is the appearance of $(\epsilon - 2)^{-1}$ poles for the purely power-law pumping \eqref{eq:power_pump}, for which, to our knowledge, no general proof of cancellation in all settings is available \cite{Vasiliev2004}.
Usually, for logarithmically divergent diagrams, such poles are compensated by the scaling of the amplitude $g_0 \nu_0^3 \sim (\epsilon - 2)$ \cite{Vasiliev2004}.
In our case, however, the relevant diagrams carry the overall factor $g_0 \nu_0$ rather than $g_0 \nu_0^3$, and this mechanism of compensation does not apply.
In any case, even if one formally drops the $(\epsilon - 2)^{-1}$ pole, the qualitative conclusion does not change: no solution with finite $c$ emerges.

\begin{figure}[h]
  \centering
  \includegraphics[width=0.75\textwidth]{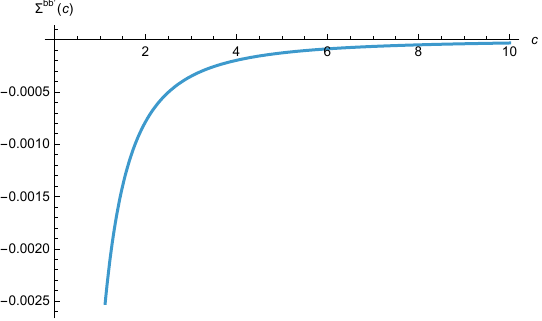}
  \caption{Graphical representation of the dimensionless function $F(c, 2)$ for the case of pumping of the type \eqref{eq:mass_pump} with parameters values  $(\mathrm{Re}, m , u) = (10^4, 1/100, 1)$ and its tendency to asymptote $F(c, 2) = 0$.}
  \label{fig:graph2}
\end{figure}

We now turn to the massive pumping \eqref{eq:mass_pump}, keeping in mind that physically realistic pumping corresponds to $\epsilon \geq 2$ \cite{Vasiliev2004}.
Substituting the massive pumping~\eqref{eq:mass_pump} into expression~\eqref{eq:ceq} and performing the momentum integration, we obtain:
\begin{equation}
\label{eq:massive_sigma}
\begin{split}
&\Sigma_{1\, \mathrm{curl}}^{b'b} = -\frac{ig_0\nu_0 m^{1-2\eps}\mathrm{Re}^{3/4}}{2c^{3}\pi^{2}(1+u_0)}\Big(c J_1(\epsilon; \mathrm{Re}, c) - J_2(\epsilon; \mathrm{Re}, c)\Big),
\end{split}
\end{equation}
where the functions $J_1$ and $J_2$ are defined as:
\begin{equation}
\begin{split}\label{eq:mass_pump_integrals}
J_1(\epsilon; \mathrm{Re}, c) &\equiv \int_{0}^{1}\mbox{d}k\, k^{2}\,(1+\mathrm{Re}^{3/{2}} k^{2})^{-\varepsilon} = \frac{\mathrm{Re}^{-3\varepsilon/2}}{3-2\varepsilon}\;
{}_2F_1\!\Big(\varepsilon,\varepsilon- {3}/{2};\varepsilon-{1}/{2};-\mathrm{Re}^{-3/{2}}\Big) \\
&+\frac{\sqrt{\pi}}{4}\,\frac{\Gamma\!\left(\varepsilon-{3}/{2}\right)}{\Gamma(\varepsilon)}\;\mathrm{Re}^{-3/2}, 
\\
J_2(\varepsilon; \mathrm{Re}, c)
&\equiv \int_{0}^{1} \mbox{d}k\, k^{3}\,(1+\mathrm{Re}^{\frac{3}{2}} k^{2})^{-\varepsilon}\,
\operatorname{arccot}\left(k/c\right), \\[6pt]
\end{split}
\end{equation}
where $\mathrm{Re} = (\Lambda/m)^{4/3}$ is the Reynolds number, all parameters in \eqref{eq:mass_pump_integrals} are assumed to be positive, and the dimensionless ansatz
$B = c\, \sqrt{u_0}\,\nu_0\Lambda$ has already been applied. 
As regards the equation $\Sigma_{1\, \mathrm{curl}}^{b'b} = 0$, the situation is the same as in the pure power-law case, in particular for $\epsilon \geqslant 2$.
The behavior of the dimensionless function $F(c, \epsilon)$ for this case is illustrated in Fig.~\ref{fig:graph2} for the parameter set $\mathrm{Re} = 10^{4}$, $m = 1/100$, $u = 1$, $\epsilon = 2$.

However, even the massive pumping \eqref{eq:mass_pump} yields only the singular solution $c = \infty$, which at first sight appears to rule out stabilization by a homogeneous mean magnetic field $\mB_0$.
The issue, however, admits a different perspective.
In the MSRDJ formalism, the pumping function, as already noted for turbulence-related problems, is a free parameter.
One might therefore assume that an inappropriate choice of pumping was made, which prevents the cancellation of the curl term.
However, inspection of the original integral (Eq.~\eqref{eq:ceq}) shows that the integrand there can be written as $I(k) = J(k)N(k)$, where $J(k) > 0$ for all $k$ in its domain and $N(k)$ is the scalar pumping function.
In principle, one could choose a sign-changing scalar pumping $N(k)$ in such a way that the curl contribution is cancelled exactly.
However, the sign-changing character of $N(k)$ leads to the loss of positive definiteness of the covariance operator of the random force in Eq.~\eqref{eq:force_correlator}.
Hence, the problem cannot be resolved merely by modifying the shape of the pumping.

It is also worth emphasizing that the curl-term problem is not intrinsically related to ultraviolet renormalization or to the renormalization-group procedure.
Indeed, in the expression for $\Sigma^{b'b}$ nothing prevents us from choosing scalar pumpings that are both UV- and IR-finite; therefore, the absence of a finite solution for $B_0$ must have a different origin. Several possible explanations can be envisaged. 
The original equations~\eqref{Eq:MHD_Eq} were formulated with stochastic pumping acting only on the velocity field, so it is conceivable that velocity fluctuations alone are insufficient to stabilize the system and that one must consider the full stochastic problem (see Ref.~\cite{Vasiliev2004}), including velocity, magnetic, and mixed noise terms.
Such an extension is, in principle, feasible; nevertheless, the prospect that adding extra noise terms will substantially cancel the curl contribution is slim.

There is, however, an alternative way out of this impasse.
In the equation for $c$ (Eq.~\eqref{eq:pure_c_eq}), if one adds a seed term of the form $i g_0\alpha$ to the right-hand side, then (as can be seen from the Figs. \ref{fig:graph1} and \ref{fig:graph2}) for a suitable choice of $\alpha > 0$ the equation admits a solution with finite $c$, and hence a finite, nonzero $B_0$. 
The question remains, however, what mechanism would generate such a seed term, since at first sight no explicit curl seed is present in the original equations.

The original MHD equations (Eq.~\eqref{Eq:MHD_Eq}) do not contain any terms responsible for the violation of spatial parity. 
This is precisely where the core of the problem lies.
By introducing parity violation in the pumping function, we effectively transmit this symmetry breaking to the entire system.
From symmetry considerations, this implies that the original MHD equations must be extended to include all terms compatible with the reduced symmetry, since parity-violating contributions are no longer forbidden.

As a consequence, Ohm’s law for a conducting fluid must be modified and reads
\begin{equation}
\label{eq:Ohm's_parity}
\mj = \sigma\big(\mE + \left[\mv\times\mB\right]/c\big)+\xi\mB,
\end{equation}
where $\xi$ is a pseudoscalar quantity.
The additional term, absent in the parity-invariant formulation, generates the required curl contribution in the derivation of the MHD equations. 
The coefficient in front of this term should be understood as an effective one, since it has no direct analogue at the microscopic level within our minimal MHD closure.
Note that a two-fluid derivation of MHD equations may generate a Biermann-battery-type term; however, such a contribution is not compatible with the constant-density approximation adopted here.

At this point, the essential feature of the MSRDJ formalism becomes manifest: it provides a description of the stationary state of fully developed turbulence. By introducing parity violation at the level of the pumping, the system is driven away from the state with 
$\langle \mb \rangle = {\bm 0}$.
During the subsequent evolution, a new term appears in Ohm’s law with a coefficient that dynamically develops as the system relaxes towards its stationary state characterized by $\langle \mb \rangle = \mB_0$.
In this stationary regime, the coefficient is replaced by a finite effective constant, which multiplies the curl term in the resulting equations.

The considerations above clarify the physical origin of the curl contribution.
Since the original MHD equations are parity invariant, the appearance of parity-violating terms must be traced back to the pumping, where spatial parity is explicitly broken.
Once this symmetry is reduced, consistency requires that the effective MHD equations include all terms allowed by the remaining symmetries.
In particular, the modified Ohm’s law necessarily generates a curl contribution with an effective pseudoscalar transport coefficient (equivalently, a pseudoscalar contribution to the conductivity).
This coefficient is not a fixed microscopic parameter but is dynamically generated as the system evolves toward a stationary turbulent state.

Thus, the appearance of curl contributions in $\Sigma^{b'b}$ does not imply a pathological inconsistency of the effective theory.
Rather, it reflects the fact that the MSRDJ formalism describes only the stationary regime of fully developed turbulence and does not capture the transient dynamics by which this regime is reached.
If the reference state with 
$\langle \mb \rangle = {\bm 0}$ becomes unstable due to these contributions, the dynamics drives the system away from this state and into a new stationary phase characterized by a nonzero magnetic field $\langle \mb \rangle = \mB_0$. 
In this stable phase, the effective action already contains a curl (seed) term whose value is fixed by the physical processes that operated before the establishment of the new stationary turbulent state. 
While the MSRDJ framework itself cannot determine this value, it does require that the system be stabilized; therefore, a seed value necessarily exists that cancels the curl contribution and yields a finite $\mB_0$ for a given ultraviolet scale $\Lambda$.

{ \section{Conclusion} \label{sec:concl} }

In this work, we revisit the issue of parity breaking in stochastic incompressible magnetohydrodynamics within the MSRDJ field-theoretic formalism, focusing in particular on the origin and physical interpretation of curl-type contributions that arise in the response functions of media with broken mirror symmetry.

By analyzing the stabilization mechanism based on a spontaneously generated uniform magnetic field $\boldsymbol{B}_0$, we show that for natural and widely used choices of the pumping function---both a pure power-law form and a massive (infrared-regularized) form---the self-consistency equation for $|\boldsymbol{B}_0|$ has only a singular solution.
In principle, such an outcome---an infinite expectation value of an unrenormalized composite field---is not unusual in quantum field theory.
In the present setting, the new putative stabilized state characterized by the one-point function $\langle\boldsymbol{b}\rangle = \boldsymbol{B}_0$ can be viewed as an analogue of the quantum-mechanical ground state (whose ground-state energy is likewise infinite and requires regularization) or of an absorbing state in percolation theory (in the sense that the system cannot be stably considered ``above'' or ``below'' this point, as in the conventional phase-transition framework).
Thus, the unregularized result should be understood merely as indicating that, for the renormalization procedure to be well defined---after which, as shown in \cite{Hnatic2025}, it acquires a finite value---an additional regularization must be imposed for the bare field $\boldsymbol{B}_0$.

We demonstrate that, in order for the self-consistency equation determining $| \boldsymbol{B}_0|$ and compensating the curl contributions to possess a finite solution (i.e., a finite uniform magnetic field), one must include bare curl terms from the outset.
Our analysis shows that such parity-violating contributions are symmetry-allowed already at the level of a generalized Ohm’s law, as the term with pseudoscalar conductivity.
Since this pseudoscalar conductivity is not fundamental at the macroscopic level (note that at the quantum level, in so-called chiral MHD describing systems with an imbalance of left- and right-handed fermions, it has an analogue in the chiral chemical potential $\mu_{5}$; see, e.g., \cite{Fukushima2008}), we treat it as an effective small parameter (for convenience, of first order in the perturbative expansion) that may be generated dynamically upon entering the fully developed turbulent regime—for instance, as a contribution to the turbulent electromotive force of $\alpha$-effect type.
Ultimately, the effective MHD equations with such an Ohm's law already contain a finite bare curl term, which provides a physical stabilization mechanism by generating a finite $\boldsymbol{B}_0$.
We emphasize that a dependence on the regularization of a nonuniversal (from the RG viewpoint) parameter such as $\boldsymbol{B}_0$ is likewise not unexpected.

Our analysis points to two consistent directions for further research.
The first is the construction of an extended stochastic MHD model that includes magnetic and mixed noise sectors, which may modify the structure of the effective action and the self-consistency conditions.
The second promising direction is a study of the $\alpha$-effect in such systems, which can be represented by the composite field $\boldsymbol{v}(\boldsymbol{x})\times\boldsymbol{b}(\boldsymbol{x})$ whose expectation value is experimentally measurable \cite{Rahbarnia2012}.
Its scaling properties can then be analyzed systematically using standard renormalization-group methods.
Both directions are necessary for a comprehensive understanding of background-field generation, helicity effects, and universal scaling behavior in turbulent magnetohydrodynamics.

\section*{Acknowledgements}
The authors would like to express their sincere gratitude to Loran Ts. Adzhemyan for valuable discussions.
The work was supported by VEGA Grant \textnumero 1/0297/25 of the Ministry of Education, Science, Research and Sport of the Slovak Republic.

\section*{Conflict of Interest:} 
The authors declare no conflicts of interest.
\appendix
\section{Appendix}\label{sec:Appendix}
In this Appendix, we present the one-loop approximation for the linear in the external momentum $\boldsymbol{k}$ contributions of the one-loop 1PI diagrams of $\Gamma^{b'b}$.
The diagrammatic contributions to diagrams depicted in \eqref{eq:one-loop_diagrams_after_shift} have the form $\rho[i\varepsilon_{ijm}k_{m}]I_{n}$, where the integrals $I_{n}$, $n = 1,2,3,4$ are listed below
\begin{equation}
\begin{split}
&I_1 = I_2 =
 \frac{i g_0 \nu _0}{48 \pi ^2 B_0^3 \left(u_0+1\right){}^2}\int\limits_0^{\Lambda}dk\frac{ N(k)}{k}\Big(-3 B_0 k^2 \nu _0^2 u_0 \left(3 u_0+1\right) \\
&+3 k \nu _0 \sqrt{u_0} \left(B_0^2 \left(u_0-1\right)+k^2 \nu _0^2 u_0 \left(3 u_0+1\right)\right) \cot ^{-1}\left(\frac{k \nu _0
   \sqrt{u_0}}{B_0}\right)+4 B_0^3\Big),
\end{split}
\end{equation}
\begin{equation}
\begin{split}
&I_3 = 
\frac{i g_0 \nu _0 }{48 \pi ^2 B_0^3  \left(u_0+1\right){}^2}\int\limits_0^{\Lambda}dk \frac{N(k)}{k}\Big(-3 B_0 k^2 \nu _0^2 u_0 \left(3 u_0+5\right)\\
&+3 k \nu _0 \sqrt{u_0} \left(B_0^2 \left(u_0+3\right)+k^2 \nu _0^2 u_0 \left(3 u_0+5\right)\right) \cot ^{-1}\left(\frac{k \nu _0
   \sqrt{u_0}}{B_0}\right)-4 B_0^3\Big),
\end{split}
\end{equation}
\begin{equation}
\begin{split}
&I_4=
-\frac{i g_0 \nu _0 }{48 \pi ^2 B_0^3  \left(u_0+1\right){}^2}\int\limits_0^{\Lambda}dk\frac{N(k)}{k}\Big(-3 B_0 k^2 \nu _0^2 \left(u_0-1\right) u_0 \\
&+3 k \nu _0 \sqrt{u_0} \left(B_0^2 \left(3 u_0+1\right)+k^2 \nu _0^2 \left(u_0-1\right) u_0\right) \cot ^{-1}\left(\frac{k \nu _0
   \sqrt{u_0}}{B_0}\right)+4 B_0^3\Big).
\end{split}
\end{equation}
Here $N(k)$ denotes the general scalar pumping function, which may depend on parameters such as $\epsilon$ or $m$. 
The integrals originating from diagrams 1 and 2 are the same.

\end{document}